\documentclass[a4paper, 11pt]{article}

\usepackage[utf8]{inputenc}
\usepackage[T1]{fontenc}
\usepackage[english]{babel}
\usepackage[pdfpagelabels]{hyperref}
\usepackage{xifthen}
\usepackage{xspace}
\usepackage{xstring}
\usepackage{calc}
\usepackage{geometry}
\usepackage{enumitem}
\usepackage{graphicx}
\usepackage[skip = 0pt]{caption}
\usepackage[affil-it]{authblk}
\usepackage{xcolor}
\usepackage{math}
\usepackage{lrmath}
\usepackage{tikz}
\usetikzlibrary{calc, external}
\usepackage{pgfplots}
\usepackage{cleveref}

\makeatletter
\newlength\@originalparskip
\newlength\@originalparindent
\renewcommand\indent{\makeatletter\hspace*{\@originalparindent}\makeatother}
\makeatother

\tikzsetexternalprefix{./figures/externalized/}
\newlength\pgfwidth
\newlength\pgfheight
\pgfplotsset{
	compat = newest,
	scale only axis,
	width = \pgfwidth,
	height = \pgfheight,
	every axis x label/.append style = {
		at = {(axis description cs:1,0)},
		anchor = north east,
		yshift = -2.8ex
	},
	every x tick label/.append style = {
		anchor = south,
		yshift = -3.3ex
	},
	every axis y label/.append style = {
		at = {(axis description cs:0,1)},
		rotate = -90,
		anchor = south west,
		yshift = +0.4ex
	},
	every y tick label/.append style = {
		anchor = east
	},
	log base 10 number format code/.code = {$\pgfmathparse{#1}\ifthenelse{\equal{\pgfmathresult}{0.0}}{1}{\ifthenelse{\equal{\pgfmathresult}{1.0}}{10}{10^{\pgfmathprintnumber\pgfmathresult}}}$},
	reverse legend,
	legend columns = 1,
	legend image code/.code = {
		\draw[mark repeat = 2, mark phase = 2] plot coordinates {
			(0cm,0cm)
			(0.2cm,0cm)
			(0.4cm,0cm)
		};
	},
	every axis legend/.append style = {
		draw = none,
		fill = none,
		legend cell align = left,
		at = {(axis description cs:1,1)},
		anchor = north east
	},
	every axis title/.style = {
		at = {(axis description cs:1,1)},
		anchor = south east,
		yshift = +0.4ex
	}
}

\title{\textbf{Critical percolation in the slow cooling of the bi-dimensional ferromagnetic Ising model}}
\date{\today}
\author{Hugo \textsc{Ricateau}}
\author{Leticia F. \textsc{Cugliandolo}}
\author{Marco \textsc{Picco}}
\affil{Sorbonne Universit\'es, Universit\'e Pierre et Marie Curie,\\ Laboratoire de Physique Th\'eorique et Hautes \'Energies,\\ 4~Place Jussieu, 75005~Paris, France}
\date{\today}

\begin{document}
	\maketitle
	\begin{abstract}
		\noindent We study, with numerical methods, the fractal properties of the domain walls found in slow quenches of the kinetic Ising model to its critical temperature.
 		We show that the equilibrium interfaces in the disordered phase have critical percolation fractal dimension over a wide range of length scales.
 		We confirm that the system falls out of equilibrium at a temperature that depends on the cooling rate as predicted by the Kibble~-- Zurek argument and we prove that the dynamic growing length once the cooling reaches the critical point satisfies the same scaling.
 		We determine the dynamic scaling properties of the interface winding angle variance and we show that the crossover between critical Ising and critical percolation properties is determined by the growing length reached when the system fell out of equilibrium.
	\end{abstract}
	\section*{Introduction}
		In recent years, the interplay between percolation and coarsening~\cite{CorberiPoliti,Bray94,HenkelPleimling10} in bi-dimensional spin models was studied in quite some detail.
		A series of papers proved that the critical and sub-critical instantaneous quench dynamics of the $2d$ ferromagnetic Ising model rather quickly approach a critical percolation state (in a time-scale that scales typically, as a small power of the system size) and later undergo the expected coarsening phenomenon that progressively makes the short length scales acquire the properties of the equilibrium target state.
		More precisely, in the quenches performed the evolution starts from a totally random initial configuration mimicking equilibrium at infinite temperature and later evolve with different microscopic stochastic spin updates.
		This feature was demonstrated with extensive numerical simulations of the Glauber~--Ising model for ferromagnetism~\cite{ArBrCuSi07,SiArBrCu07,BlCoCuPi14,BlCuPiTa17} and the Kawasaki model for phase separation~\cite{SiSaArBrCu09,TaCuPi16} quenched into their symmetry broken phases.
		The effects of weak disorder were considered in~\cite{SiArBrCu08,InCoCuPi16}.
		The voter model dynamics was investigated in~\cite{TaCuPi15} and, especially relevant for the present study, quenches to the critical point of the $2d$ ferromagnetic Ising model were considered in~\cite{BlCuPi12}.
		The early approach to critical percolation also explained why zero temperature quenches of the $2d$ Ising model often get blocked in metastable states with infinitely long-lived flat interfaces~\cite{SpKrRe01,SpKrRe02,BaKrRe09,OlKrRe12,OlKrRe11a,OlKrRe11b,OlKrRe13}.
		Metastable states in quenches from the critical point to zero temperature were considered in~\cite{BlPi13}.
		
		In statistical physics studies, quenches are taken to be instantaneous.
		Indeed, the relevant time scales in experimental realisations are such that the cooling time is much shorter than all other time scales.
		Instead, in field theoretical models of cosmology, there was interest in determining the cooling rate dependencies induced by a very slow quench across a second order phase transition.
		The original Kibble arguments for the existence of spatial regions that are not causally connected long after going through the phase transition~\cite{Kibble76} were complemented by a scaling proposal by Zurek~\cite{Zurek85,Zurek96}.
		This argument allows one to estimate the correlation length reached when the system falls out of equilibrium approaching a critical point from the symmetric phase with a weak finite speed.
		The interest in counting the number of topological defects left over after crossing the phase transition triggered by cosmology~\cite{Kibble07}, prompted condensed-matter experimental physicists to try these measurements in the lab.
		This kind of experiments were first performed in Helium 3~\cite{Bauerle-etal} and liquid crystals~\cite{Chuang-etal} more than twenty years ago.
		The subject was recently revived by the realisation of cold atom experiments in which the samples are taken across the critical region with a finite speed~\cite{Lamporesi-etal,Chomaz-etal,Navon-etal,Donatello-etal}.
		New studies in ion crystals~\cite{delCampo-etal,Ulm-etal13,Silvi-etal16} and $2d$ colloidal suspensions~\cite{Maret} have also been recently performed.
		Two recent reviews give a more complete summary of the status of this field~\cite{delCampo10,Beugnon17}.
		
		Studies of cooling rate dependencies in statistical physics models were performed in a number of papers.
		The $2d$ Ising model with non-conserved order parameter dynamics was considered in~\cite{Biroli10} and the $2d$ xy (planar spins) model in~\cite{Jelic11} (the latter is relevant to discuss the recent experimental activity in Bose~-- Einstein condensates and colloidal suspensions).
		In the former model the phase transition is a conventional second order one, from a symmetric to a symmetry broken phase, while in the latter case the transition is of Berezinskii~-- Kosterlitz~-- Thouless (BKT) kind and the target phase is a critical one.
		The aim of these papers was to show that, contrary to what was usually claimed in the KZ literature, the dynamics are not frozen after the system falls out of equilibrium close to the critical point, be it second order or BKT.
		The critical or subcritical dynamics, at continuously changing control parameters, let the dynamic correlation length go on growing in time.
		Scaling arguments were used in these papers to derive the dependence of the growing correlation length, and hence the number of topological defects, as a function of time and cooling rate and they were favourably compared to the outcome of numerical simulations.
		Exact results for the one dimensional Ising chain and a variety of cooling procedures were derived in~\cite{Krapivsky10}.
		The spherical ferromagnetic model with exponentially fast cooling was treated, also analytically, in~\cite{Picone03}.
		A one-dimensional non-equilibrium lattice gas model with a phase transition was treated in~\cite{Priyanka}.
		Extensive numerical simulations of models for two dimensional atomic gases were very recently presented in~\cite{Comaron17a,Comaron17b}.
		The evolution of the order parameter in the finite dimensional Ising model slowly cooled to the critical point were studied with different microscopic stochastic rules in~\cite{Liu-etal14}.
		
		In this paper we revisit the slow cooling of the $2d$ Ising model~\cite{Biroli10,Liu-etal14} paying now special attention to the geometric properties of the domain structures formed when approaching the critical point.
		The paper is organised as follows.
		In \cref{sec:model} we present the model and the observables.
		\Cref{sec:equilibrium} summarises some features of the equilibrium state that are useful for the dynamic study.
		We then present a short account of instantaneous quenches in \cref{sec:quenches} and we finally enter the heart of the results on the cooling rate measurements in \cref{sec.cooling}.
		The last section sums up our results.
	\section{The model and the observables}
	\label{sec:model}
		\subsection{The bi-dimensional ferromagnetic Ising model}
			We focus on the emblematic ferromagnetic Ising model
			\begin{equation*}
				H\lr{(\lr{\{\sigma_i\}})}=-J\sum_{\lr{<i,j>}}\sigma_i\,\sigma_j\eqpc
			\end{equation*}
			with $J>0$, and spin variables taking only two values, $\sigma_i=\pm1$.
			In particular, we study its bi-dimensional $d=2$ realisation on the square lattice, so that the symbol $\lr{<i,j>}$ represents a sum over nearest neighbours only.
			The total number of spins in the system is $L\times L$ with $L$ the linear length of the lattice measured in units of the lattice spacing $a$.
			The canonical equilibrium properties as a function of the parameter $K=\beta\,J$, with $\beta$ the inverse temperature, are described by the partition function
			\begin{equation*}
				Z\lr{(K)}=\sum_{\mathclap{\lr{\{\sigma_i=\pm1\}}}}\exp{K\sum_{\lr{<i,j>}}\sigma_i\,\sigma_j}\eqpd
			\end{equation*}
			Hereafter we work with units such that $J=1$.
			
			The model is endowed with microscopic Monte Carlo stochastic dynamics for the individual spins.
			The microscopic update rule is defined as follows: we randomly chose a spin $i\in\lr{\llbracket0,L^2\llbracket}$ in the system.
			The spin is flipped ($\sigma_i=-\sigma_i$) with a probability
			\begin{equation*}
				p=\min\lr{(1,e^{-\beta\,\delta H})}\eqpc
			\end{equation*}
			where $\delta H$ is the energy change due by the potential flip of the selected spin.
			$\beta\,\delta H$ can only takes five different values: $-8K$, $-4K$, $0$, $4K$, or $8K$.
			The process is controlled by the parameter $K$ given by the external inverse temperature of the bath $\beta$ times the exchange parameter $J$.
			Repeating this process $L^2$ times constitutes one unit of time in the kinetic Ising model.
			Hereafter, the time appearing in dynamical studies is always in this unit.
			We use a square lattice with linear size $L=1024$ and periodic boundary conditions.
		\subsection{Percolation}
			Site percolation~\cite{Stauffer94,Christensen02,Saberi15,Delfino15} is a purely geometric problem in which particles are placed at the sites of a lattice with probability $p$.
			This model undergoes a phase transition at $p_{\text{c}}$, a critical value of $p$ that depends on the geometry and dimension of the lattice.
			In $d=2$, and for a square lattice, $p_{\text{c}}\sim0.59$.
			The two phases correspond to one with no cluster spanning the system from one end to the other in any spatial directions ($p<p_{\text{c}}$) and one in which there is one cluster percolating across the lattice ($p>p_{\text{c}}$).
			At the critical point the behaviour is similar to the one at a thermodynamic second order critical point with universal critical exponents characterising various geometric quantities that one can define.
			
			The Ising model can be thought of as a percolation problem after performing a one-to-one mapping between spins and occupation numbers.
			For example, an infinite temperature configuration in which the spins take $\pm$ values with probability $\nicefrac{\displaystyle{1}}{\displaystyle{2}}$ is a random percolation configuration with $p=\nicefrac{\displaystyle{1}}{\displaystyle{2}}$.
			It is, therefore, below the threshold for percolation of a cluster of occupied sites on the square lattice.
		\subsection{Critical behaviour: fractal domains}
			We will investigate the properties of geometric domains in the kinetic Ising model, that is to say, ensembles of connected spins pointing in the same direction (surrounded by a domain of the opposite orientation when in the bulk, or reaching the boundaries of the system if open boundary conditions are used).
			
			At the critical point of the $2d$ ferromagnetic Ising model the geometric domains are fractal objects.
			Their typical area and typical interface length are
			\begin{equation*}
				A_{\text{c}}=\ell^{D_A}\eqpc\qquad\qquad\ell_{\text{c}}=\ell^{D_\ell}\eqpc
			\end{equation*}
			with $\ell$ a typical length.
			The Hausdorff dimensions are given by
			\begin{equation*}
				D_A=1+\frac{3\,\kappa}{32}+\frac{2}{\kappa}\eqpc\qquad\qquad D_\ell=1+\frac{\kappa}{8}\eqpc
			\end{equation*}
			with $\kappa$ a universal parameter that characterises the critical point.
			At the thermodynamic critical point of the Ising model in dimension two, $\kappa=3$.
			Instead, at the percolation threshold $\kappa=6$.
			In this study we only analyse interface properties so we will only use $D_\ell$ in the rest of the paper.
		\subsection{Observables}
			We used a small number of observables that are enough to characterise the growing length and geometric properties of the interfaces.
			We define them in this section.
			\subsubsection{Space time correlation function and correlation length}
				In equilibrium, the correlation of the spin fluctuations
				\begin{equation}
					C_{\text{c}}\lr{(r=\lr{|i-j|})}=\lr{<\sigma_i\,\sigma_j>}-\lr{<\sigma_i>}\,\lr{<\sigma_j>}\eqpc
					\label{eq:def-corr}
				\end{equation}
				where $r\in\lr{\llbracket0,\nicefrac{\displaystyle{L}}{\displaystyle{2}}\rrbracket}$, allows one to extract the equilibrium correlation length $\xi_{\text{eq}}$ with different studies of its decaying properties over distance.
				For example, one can use a fit to the expected form close to the critical point,
				\begin{equation*}
					C_{\text{c}}\lr{(r)}=\frac{e^{\nicefrac{\scriptstyle{-r}}{\scriptstyle{\xi}_{\text{eq}}}}}{r^{a}}\eqpc
				\end{equation*}
				or extract it from the weighted integral
				\begin{equation}
					\xi_{\text{eq}}=\frac{\displaystyle{\Gamma\lr{(\zeta-a)}\,\int_0^\Lambda r^\zeta\,C_{\text{c}}\lr{(r)}\,\d{r}}}{\displaystyle{\Gamma\lr{(\zeta-a+1)}\,\int_0^\Lambda r^{\zeta-1}\,C_{\text{c}}\lr{(r)}\,\d{r}}}
					\label{eq:corr-length}
				\end{equation}
				with a convenient choice of the power $\zeta$ and the cut-off length $\Lambda$.
				In particular, we will use $\zeta=2$ and $\Lambda$ is chosen the largest possible such that $C_{\text{c}}\lr{(r)}$ remains larger than its statistical fluctuations.
				
				In dynamical studies, the space-time correlation is defined just as in eq.~(\ref{eq:def-corr}) where the spins are time-dependent variables.
				The average is taken over different histories (random noises) of the dynamics.
				After a quench, while the system is far from equilibrium, the spin average vanishes and the connected and plain correlations simply coincide.
				The procedure in the right-hand-side of eq.~(\ref{eq:corr-length}) can then be applied to extract the dynamic growing length $\xi(t)$ that characterises the growth of equilibrium structures close, at, and below $T_{\text{c}}$.
			\subsubsection{Variance of the interfaces winding angle}
				The winding angle, $\theta\lr{(\ell)}$, is measured on any bi-dimensional curve as a function of the curvilinear abscissa, $\ell$, as follows.
				We first choose an origin point for the curvilinear abscissa.
				Then, we measure a reference angle, $\theta_0$, between a chosen fixed direction and the tangent to the curve at the origin of the curvilinear lengths.
				Now, for each point on the curve, we define $\eta\lr{(\ell)}$, the local angle between the same chosen fixed direction as earlier and the tangent to the curve at $\ell$.
				Finally, the winding angle is obtained by integrating the variation of the local angle along the curve:
				\begin{equation}
					\theta\lr{(\ell)}=\theta_0+\int_0^\ell\d{\eta}
					\label{eq:wav-equil}
				\end{equation}
				(note that on a square lattice $\eta$ can only take four values).
				For closed curves, after one turn (\emph{ie} returning to the origin), we have $\Delta\theta=2n\pi$, where $n\in\mathbb{Z}$ is the number of loops.
				In particular, since the curve is an interface, it cannot cross itself and $\Delta\theta=0$ or $\Delta\theta=\pm2\pi$ (where the sign changes whether the curve rotates clockwise or anticlockwise).
				The former means that the interface encloses a finite area, and the latter means that the interface spans the system from one border to another one.
				
				The moments of these angles can then be computed by taking their desired power and performing the equilibrium or dynamic statistical averages.
				
				For a fractal curve the average of the angle vanishes and its variance satisfies~\cite{SaDu87}
				\begin{equation*}
					\lr{<\theta^2\lr{(\ell)}>}=C+\frac{4\,\kappa}{8+\kappa}\,\log{\ell}\eqpc
				\end{equation*}
				where $\ell$ is the curvilinear distance along the curve, $C$ is a non-universal constant, and $\kappa$ takes a universal value depending on the kind of criticality.
				
				In the dynamic model, this definition can be applied to study the evolution of the geometric properties of the interfaces in the system.
	\section{Equilibrium behaviour}
		\label{sec:equilibrium}
		In this section we review some properties of the equilibrium behaviour of the $2d$ Ising model at high temperature and at the critical point that are relevant to our study.
		\begin{figure}[!htb]
			\begin{center}
				\includegraphics{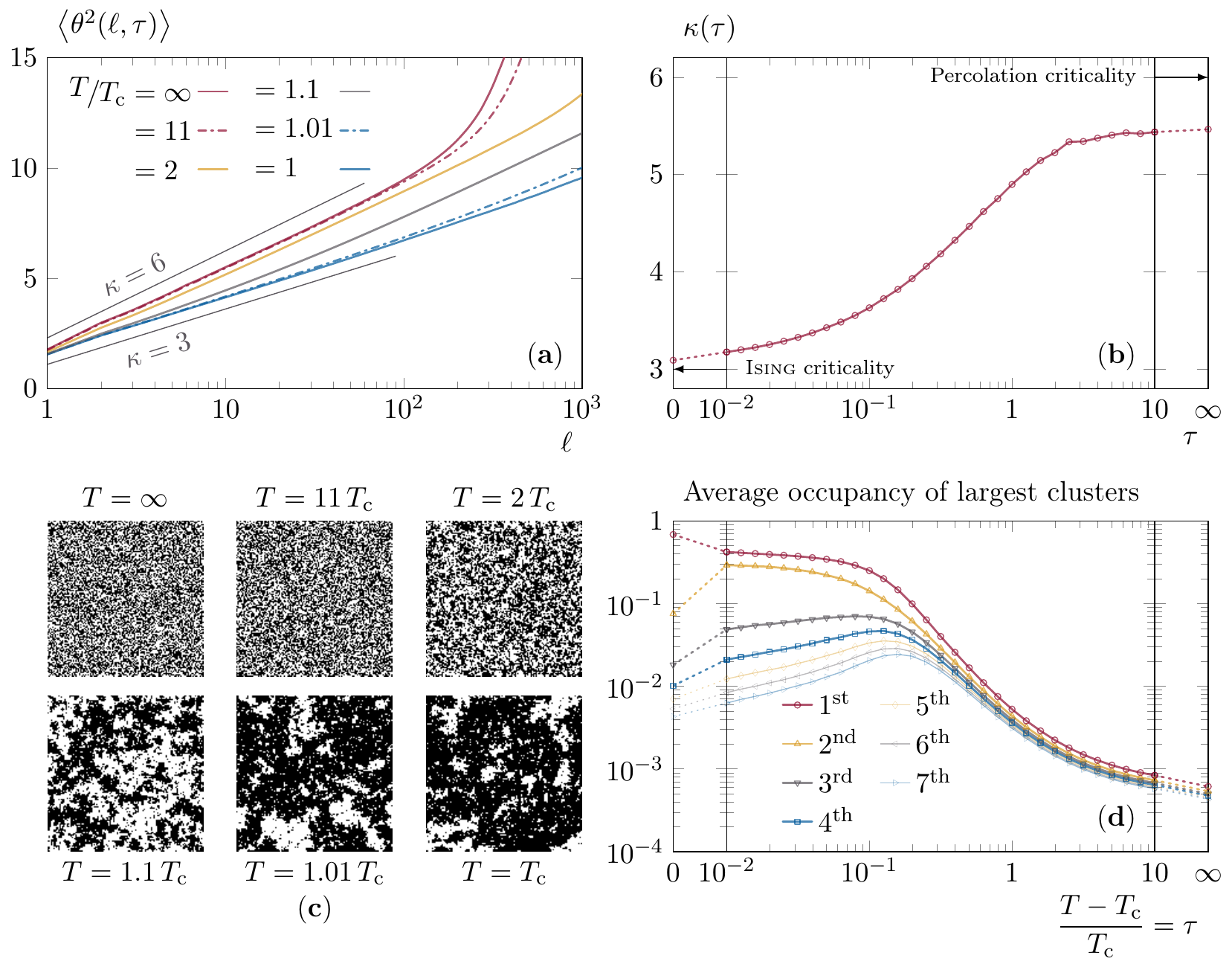}
			\end{center}
			\caption{
				Equilibrium behaviour above the Curie point ($T\geq T_{\text{c}}$).
				Panel (\textbf{a}) shows the winding angle variance as a function of the curvilinear length on the interfaces, at different temperatures.
				The two straight lines, $\kappa=3$ and $\kappa=6$, are the expected slopes for the Ising and percolation universality classes, respectively.
				Panel (\textbf{b}) displays, as a function of $T$, the value of $\kappa$ extracted from the slope of $\left<\theta^2\left(\log{\ell}\right)\right>$ at short length $\ell$.
				The horizontal axis, the same as on the graphic below, is a logarithmic scale where we added the two extreme points, $0$ and $\infty$.
				The values of $\kappa$ corresponding to the two universality classes (Ising and percolation) are labeled on the graph.
				The lower row shows, in (\textbf{c}), typical snapshots of the equilibrium state of the system ($L=128$), at different temperatures above $T_{\text{c}}$, and in (\textbf{d}) the average occupancy rates of the first largest clusters when approaching the critical temperature.
				The $n^{\text{th}}$ average occupancy is the fraction of the system occupied, on average, by the $n^{\text{th}}$ largest cluster.
				Except for the snapshots, all the results presented in this figure were obtained using $L=1024$.
			}
			\label{equilibrium.plots}
		\end{figure}
		
		We start by recalling a number of thermal features above the critical temperature.
		With this, we want to establish a reference equilibrium behaviour for the relevant observables.
		
		Away from the critical point correlations span a finite distance.
		The equilibrium correlation length diverges at the Curie critical temperature, and in the close vicinity of the critical point, it does as a power law,
		\begin{equation}
			\xi_{\text{eq}}\lr{(\tau)}\sim\tau^{-\nu}\qquad\text{where}\qquad\tau=\frac{T-T_{\text{c}}}{T_{\text{c}}}>0
			\label{equilibrium.xi.taunu}
		\end{equation}
		is the distance to the critical point\footnote{here we are only interested in the behaviour above the Curie temperature (\emph{ie} $T>T_{\text{c}}$).}, and $\nu=1$ is the universal critical exponent of the Ising universality class associated to the correlation length.
		\Cref{equilibrium.xi.taunu} is only valid in a close vicinity of the critical temperature ($\tau\ll1$); far from it, there are extra corrections to add, but we do not need them here.
		Another limitation of \cref{equilibrium.xi.taunu} is that it is only valid for an infinite system; if the system size ($L$) is finite, it limits the growth of the correlation length to a saturation threshold that scales with the system size as $\xi_{\text{eq}}\lr{(\tau=0)}=\bar{\xi}_{\text{eq}}\sim L$.
		
		Let us now discuss the equilibrium behaviour of the variance of the winding angle (\textsc{wav}), \emph{ie} the nature of the interfaces between domains, see \cref{equilibrium.plots}~(\textbf{a}).
		We observe that the \textsc{wav} increases logarithmically on short curvilinear length scales; the value of $\kappa$ extracted from the slope of $\lr{<\theta^2\lr{(\log\ell)}>}$ is close to $6$ at high temperature and close to $3$ at $T_{\text{c}}$.
		This means that, on short length scales, the interfaces of the domains are subject to a conformal invariance (with the criticality of percolation at high temperature and the one of Ising at $T_{\text{c}}$).
		There is nothing surprising here.
		Firstly, at the Ising critical point the domains obviously have the criticality of the corresponding universality class.
		Secondly, at high temperature, the Ising model is a percolation problem (correlations are so short that one could argue that the spins are randomly chosen to point up or down with half probability, $p=\nicefrac{\displaystyle{1}}{\displaystyle{2}}$).
		A typical configuration is, therefore, one of a site percolation problem away from its critical point (recall that, on a square lattice, the critical percolation threshold is at $p_{\text{c}}\approx0.593>\nicefrac{\displaystyle{1}}{\displaystyle{2}}=p$).
		In consequence, on average, there are no percolating clusters in these configurations.
		This means that the conformal invariance disappears at sufficiently long length scales: $\ell\sim{\lr{|p-p_{\text{c}}|}}^{-\nu_{\text{p}}D_{\ell}}\sim10^2$, where $\nu_{\text{p}}=\nicefrac{\displaystyle{4}}{\displaystyle{3}}$ is the percolation correlation length critical exponent and $D_{\ell}=\nicefrac{\displaystyle{7}}{\displaystyle{4}}$ is the fractal dimension of the interface of a percolation cluster.
		This leads us to our second remark: at high temperature and long length scales, the \textsc{wav} does not grow logarithmically anymore; it increases much faster.
		This is, in fact, due to the finite size of the domains.
		Indeed, since we are far from the critical percolation threshold, the domains remain small, and the overall curvature necessary to close their interface is responsible for a faster growth of the \textsc{wav}.
		When the temperature decreases the domains swell (like the correlation length), and the \textsc{wav} stops its logarithmic growth at a longer and longer length scale.
		Obviously, when reaching $T_{\text{c}}$, there is a true conformal invariance, and the \textsc{wav} increases logarithmically on all length scales.
		Considering only the short length scales, as the temperature decreases, the criticality smoothly evolves from the percolation universality class to the Ising one.
		This is most clearly shown in panel (\textbf{b}) in \cref{equilibrium.plots} where $\kappa$ is plotted as a function of $T$.
		The slope is extracted from the \textsc{wav} by linear interpolation on short length scales; the longer length scales, where criticality disappears, are excluded from the interpolation set.
		The Ising criticality is only reached in a close vicinity of the critical point ($T<1.1\,T_{\text{c}}$).
		
		The fact that we observe critical percolation properties in the disordered phase is related to the presence of a critical curve in the temperature-field phase diagram of the $2d$ Ising model.
		It separates a phase with an infinite cluster of parallel spins (at sufficiently large external field) from one without (weak field).
		This critical curve joins the Ising critical point (Curie temperature and zero field) with the infinite temperature limit at non-vanishing value of the external field, while remaining close to the zero field axis~\cite{Delfino15}.
		The vicinity of this line at our working temperatures justifies the fact that we see (finite size) critical percolation geometric properties on the spins clusters.
		
		The last quantity we want to discuss is the average occupancy rate of the largest clusters shown in \cref{equilibrium.plots} (\textbf{c}).
		Firstly, at high temperature, all the clusters are more or less of the same size.
		Then, when temperature decreases, the bigger clusters start to grow by absorbing the smaller ones, up to a point ($T\approx1.1\,T_{\text{c}}$) where only the two biggest prevail over all the others.
		Having two coexisting big clusters is a feature of percolation of Ising clusters\footnote{\label{footnote.equilibrium.percoisingclusters} in site percolation, at $p=p_{\text{c}}$, the largest cluster (the percolating one) is much larger than the second one (of approximately one order of magnitude).
		In Ising models two percolating clusters are in competition: the up spins one and the down spins one; generally, in a $q$-state Potts model, the $q$ largest clusters are of the same order of magnitude while the $q+1$-th will be much smaller.}.
		These two clusters will coexist up to a very close vicinity of the Curie temperature ($T\lesssim1.01\,T_{\text{c}}$).
		In contrast, at the Ising critical point there is only one large cluster (much larger than all the others).
		
		To summarise, at $T_{\text{c}}$, or in its very close vicinity, the system is occupied by only one large geometric cluster having the Ising criticality ($\kappa\approx3$) at all length scales.
		See the snapshot at $T=T_{\text{c}}$ in \cref{equilibrium.plots} (\textbf{c}).
		At high temperature, the domains are much smaller.
		However, on short length scales, they have the geometric properties of critical percolation ($\kappa\approx5.5$, which is only $5\%$ different from the slope expected with $\kappa=6$).
		Finally, in between, the criticality smoothly changes from the percolation one to the Ising one in the range $\lr{[T_{\text{c}},1.1\,T_{\text{c}}]}$; the coexistence of the two biggest clusters ends much closer to the critical point ($T\lesssim1.01\,T_{\text{c}}$).
	\section{Instantaneous quenches}
	\label{sec:quenches}
		In this section we recall some features of the dynamics after instantaneous quenches to zero temperature and the critical point, as interpreted from the geometric point of view that we adopt in this paper.
		\subsection{Quench to \texorpdfstring{$T=0$}{T=0}}
			\begin{figure}[!htb]
				\begin{center}
					\includegraphics{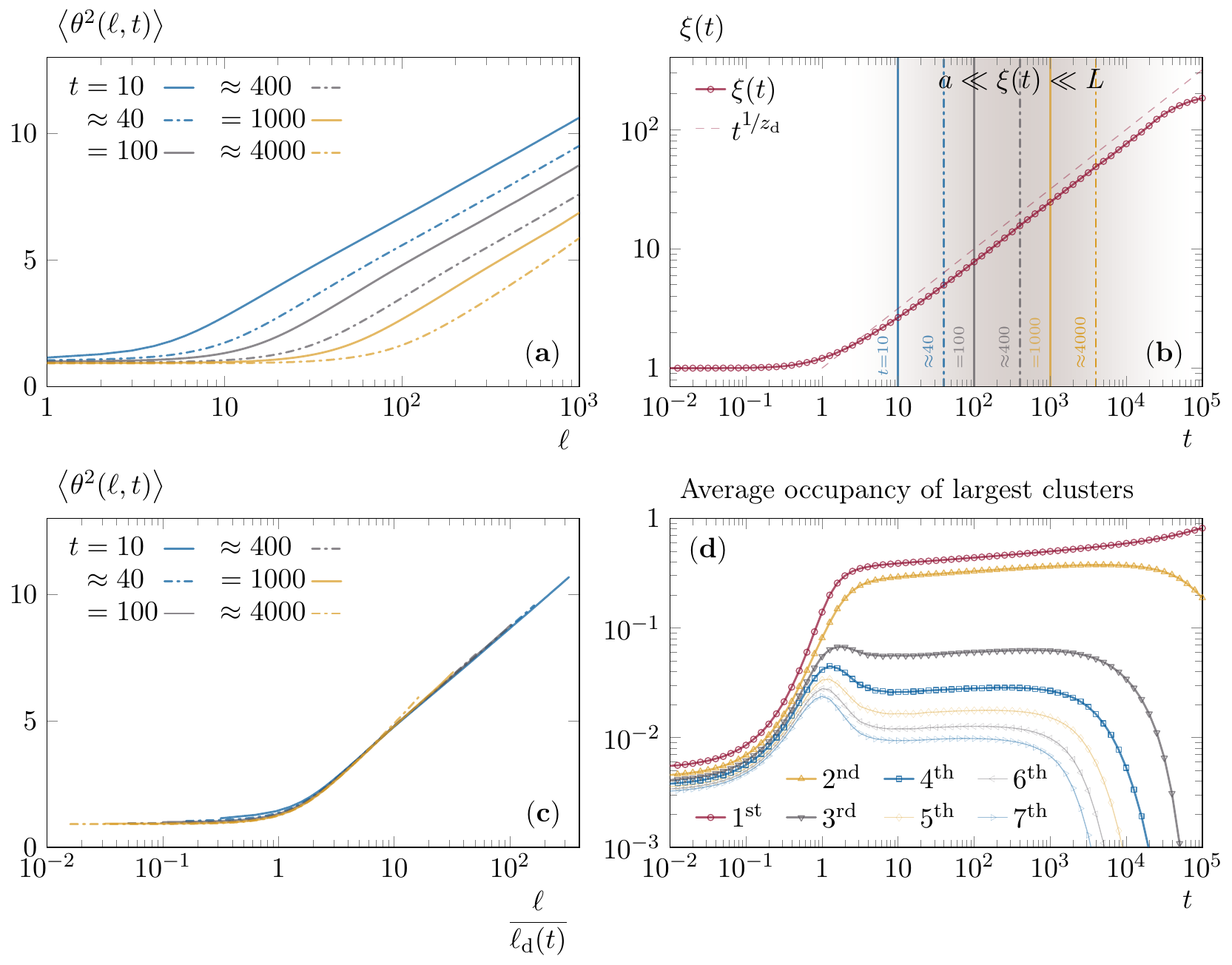}
				\end{center}
				\caption{
					Out-of-equilibrium evolution in post-quench dynamics (from $T=2\,T_{\text{c}}$ to $T=0$).
					Panel (\textbf{a}) represents the winding angle variance (\textsc{wav}) at different times following the quench.
					These times are reported in panel (\textbf{b}), and are chosen such that the constraint $a\ll\xi\left(t\right)\ll L$ is fulfilled.
					Panel (\textbf{c}) still represents the \textsc{wav}, but with a different scaling: the horizontal axis is rescaled following \cref{instaquench.0.wav.scaling}, and $\ell_{\text{d}}\left(t\right)$ is evaluated through its theoretical expression ($\sim t^{\nicefrac{\scriptstyle{1}}{\scriptstyle{z_{\text{d}}}}}$).
					Panel (\textbf{b}) shows the evolution over time of the correlation length $\xi\left(t\right)$ extracted from the space-time two point correlation function.
					Its theoretical time-dependence is shown with a dashed line; the range of validity of this prediction is highlighted by the grey shading ($a\ll\xi\left(t\right)\ll L$).
					Panel (\textbf{d}) represents, as a function of time, the average occupancy rates of the first largest clusters (see \cref{equilibrium.plots}).
					All the results presented in this figure were obtained using a system size $L=1024$.
				}
				\label{instaquench.0.plots}
			\end{figure}
			
			The second situation of interest is the one of an instantaneous quench to zero temperature.
			We consider the following procedure: starting from an equilibrium state at $T=2\,T_{\text{c}}$, at $t=0$ we suddenly change the temperature of the bath to zero, \emph{ie}
			\begin{equation*}
				T\lr{(t)}=\lr{\{\begin{array}{ll}
					2\,T_{\text{c}}&t\leq0\\
					0&t>0
				\end{array}.}
			\end{equation*}
			and we observe the further evolution of the system.
			
			In such a procedure, the growing length is known to increase as a power law,
			\begin{equation*}
				\xi\lr{(t)}\sim t^{\nicefrac{\scriptstyle{1}}{\scriptstyle{z_{\text{d}}}}}\eqpc
			\end{equation*}
			where $z_{\text{d}}=2$ is the dynamical exponent~\cite{AlCa79,Bray94}.
			Of course, this result holds only for $t$ such that $a\ll\xi\lr{(t)}\ll L$: $\xi\lr{(t)}$ cannot be smaller than the lattice spacing, and it is bounded by the finite system size.
			The growing length extracted from the correlation function after such a quench is compared to the theoretical expectation in \cref{instaquench.0.plots}~(\textbf{b}).
			
			Let us now discuss how the behaviour of the \textsc{wav} evolves in time, as displayed in \cref{instaquench.0.plots} (\textbf{a}).
			At the initial time, the system is in equilibrium, and the \textsc{wav} behaves as described in eq.~(\ref{eq:wav-equil}), see \cref{equilibrium.plots} (\textbf{a}).
			Then, the zero temperature dynamics start to smooth the interfaces: first, on short length scales, then, on longer length scales.
			This is the first part of the curve and the \textsc{wav} does not increase since a smooth interface has no criticality.
			In the meantime, the clusters swell, and since the system has not yet realised, at long length scales, that it is at zero temperature (and should have smooth interfaces), it develops the criticality of percolation.
			This is the second part of the curve; the \textsc{wav} restarts to grow logarithmically.
			The typical (curvilinear abscissa) length scale that separates these two behaviours is denoted $\ell_{\text{d}}\lr{(t)}$, and is related to the typical size of the domains:
			\begin{equation*}
				\ell_{\text{d}}\lr{(t)}\sim {\xi\lr{(t)}}^D\sim t^{\nicefrac{\scriptstyle{D}}{\scriptstyle{z_{\text{d}}}}}=\sqrt{t}\eqpc
			\end{equation*}
			since $D=1$ is the (fractal) dimension of the smooth interfaces on short length scales.
			The \textsc{wav} has a universal behaviour in time that is highlighted by the rescaling
			\begin{equation}
				%\lr{<\theta^2\lr{(\ell,t)}>}\to\lr{<\theta^2\lr{(\ell,t)}>}\qquad\text{and}\qquad
				\ell\to\frac{\ell}{\ell_{\text{d}}\lr{(t)}}\eqpc
				\label{instaquench.0.wav.scaling}
			\end{equation}
			once again, while $\xi(t)$ is in the range $a\ll\xi\lr{(t)}\ll L$.
			See \cref{instaquench.0.plots} (\textbf{c}).
			
			\Cref{instaquench.0.plots} (\textbf{d}) shows the evolution of the average size of the largest clusters.
			Starting from a high-temperature equilibrium state, all the clusters are almost of the same size.
			Next, in the early dynamics, they all grow in the same way.
			As soon as the correlation length starts to grow, the larger clusters progressively swallow the smaller ones.
			Indeed, the smaller the clusters, the faster they disappear.
			This is the so-called coarsening dynamics.
			In particular, the second largest cluster lengthly coexists with the largest one.
			As already mentioned, this long coexistence of two large clusters having almost the same size (the third cluster is far smaller) is a typical feature of percolation of Ising clusters (see \cref{footnote.equilibrium.percoisingclusters} page~\pageref{footnote.equilibrium.percoisingclusters}).
			
			In the course of this process, the quench protocol went through the Ising critical point, and there is no track of it.
			Now the question is: what happens if, like in real experiments, we cannot do the quench instantaneously?
			What is the influence of the time spent near the Curie temperature?
			\Cref{sec.cooling} will address these questions.
			However, let us first explore the dynamics after an instantaneous quench to the Curie temperature.
		\subsection{Quench to \texorpdfstring{$T=T_{\text{c}}$}{T=Tc}}
		\label{sec.quench.atTc}
			\begin{figure}[!htb]
				\begin{center}
					\includegraphics{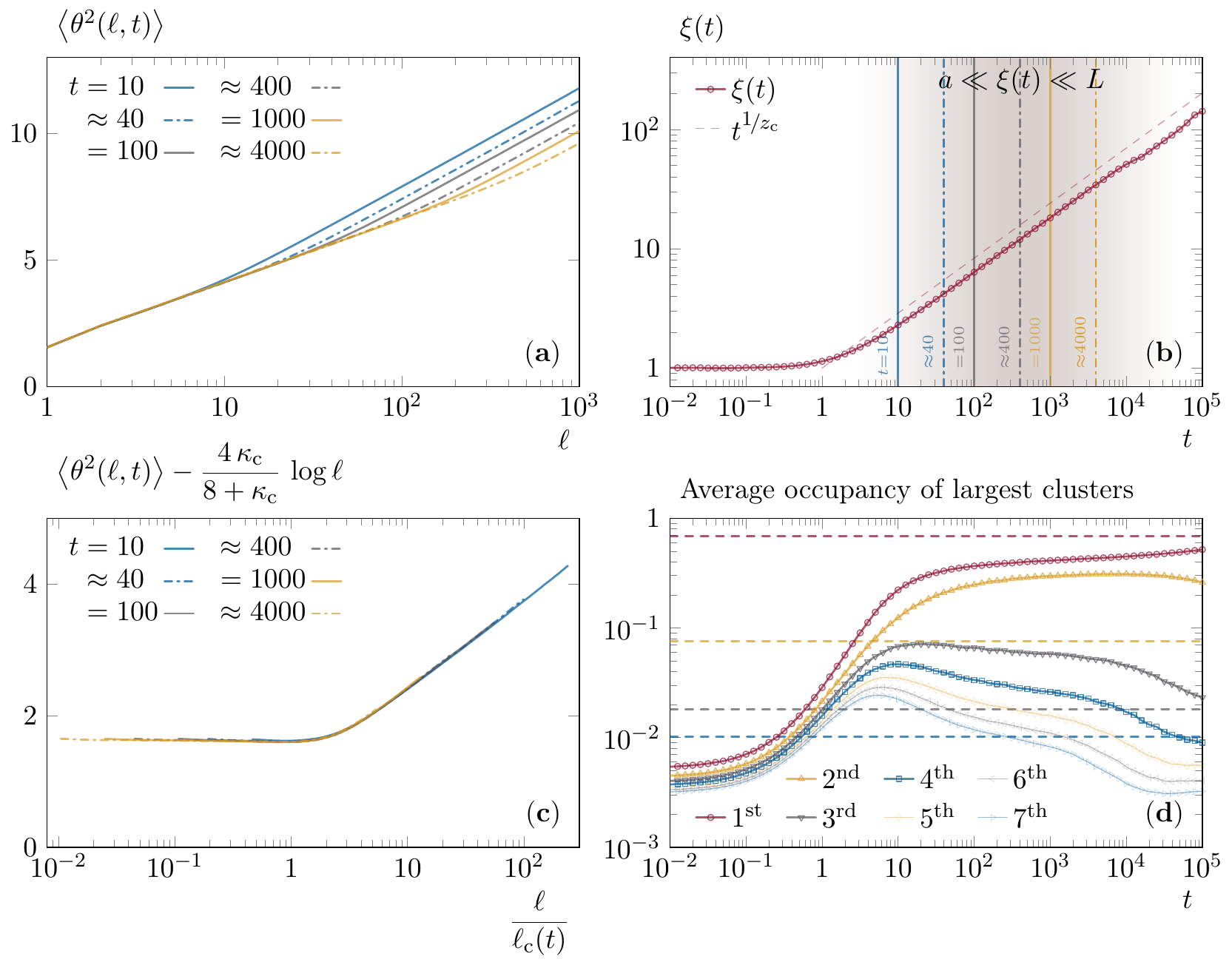}
				\end{center}
				\caption{
					Out-of-equilibrium evolution in critical post-quench dynamics (temperature is instantaneously taken from $T=2\,T_{\text{c}}$ to $T=T_{\text{c}}$).
					The graphics are organised in the same manner as in \cref{instaquench.0.plots}.
					However, panel (\textbf{c}) is now scaled following \cref{instaquench.Tc.wav.scaling}, where $\ell_{\text{c}}\left(t\right)$ is evaluated through its theoretical expression ($\sim t^{\nicefrac{\scriptstyle{D_{\text{c}}}}{\scriptstyle{z_{\text{c}}}}}$).
					Moreover, in the lower right figure, we added with dashed lines the equilibrium values of the average occupancy rates of the first largest clusters at $T_{\text{c}}$.
				}
				\label{instaquench.Tc.plots}
			\end{figure}
			
			The process is the same as above, except that the temperature immediately after $t=0$ is now the Curie temperature:
			\begin{equation*}
				T\lr{(t)}=\lr{\{\begin{array}{ll}
					2\,T_{\text{c}}&t\leq0\\
					T_{\text{c}}&t>0
				\end{array}.}
			\end{equation*}
			
			In this situation, the correlation length still grows as a power law,
			\begin{equation*}
				\xi\lr{(t)}\sim t^{\nicefrac{\scriptstyle{1}}{\scriptstyle{z_{\text{c}}}}}\eqpc
			\end{equation*}
			with $z_{\text{c}}\approx2.17$ the critical dynamical exponent~\cite{Hohenberg-Halperin,BlCuPi12,NightingaleBlote00,Duclut}.
			The growth of the correlation length is slightly slower than in the previous situation since $\nicefrac{\displaystyle{1}}{\displaystyle{z_{\text{c}}}}\approx0.461<\nicefrac{\displaystyle{1}}{\displaystyle{2}}$.
			See panel (\textbf{b}) in \cref{instaquench.Tc.plots}.
			Again, this result is only true for $\xi(t)$ in the range $a\ll\xi\lr{(t)}\ll L$.
			
			Concerning the \textsc{wav}, it behaves exactly as in the zero temperature quench except that, instead of the smooth zero temperature thermal state, it is the Ising criticality that develops over short length-scales, see \cref{instaquench.Tc.plots} (\textbf{a}).
			The typical (curvilinear abscissa) length scale that separates the Ising criticality from the percolation one, $\ell_{\text{c}}\lr{(t)}$, scales differently with the correlation length:
			\begin{equation*}
				\ell_{\text{c}}\lr{(t)}\sim {\xi\lr{(t)}}^{D_{\text{c}}}\sim t^{\nicefrac{\scriptstyle{D_{\text{c}}}}{\scriptstyle{z_{\text{c}}}}}\eqpd
			\end{equation*}
			Since the interfaces on short length scales are not smooth anymore, their fractal dimension is given by
			\begin{equation*}
				D_{\text{c}}=1+\frac{\kappa_{\text{c}}}{8}=1.375\eqpc
			\end{equation*}
			where $\kappa_{\text{c}}=3$ is the same universal parameter as in the pre-factor in front of the logarithmic growth of the \textsc{wav}.
			Note that $\nicefrac{\displaystyle{D_{\text{c}}}}{\displaystyle{z_{\text{c}}}}\approx0.634>\nicefrac{\displaystyle{1}}{\displaystyle{2}}$.
			The \textsc{wav} still has a universal behaviour, now highlighted by the rescaling
			\begin{equation}
				\lr{<\theta^2\lr{(\ell,t)}>}\to\lr{<\theta^2\lr{(\ell,t)}>}-\frac{4\,\kappa_{\text{c}}}{8+\kappa_{\text{c}}}\,\log{\ell}\qquad\text{and}\qquad\ell\to\frac{\ell}{\ell_{\text{c}}\lr{(t)}}\eqpc
				\label{instaquench.Tc.wav.scaling}
			\end{equation}
			where $\nicefrac{\displaystyle{4\,\kappa_{\text{c}}}}{\displaystyle{\lr{(8+\kappa_{\text{c}})}}}\approx1.09$, and still while $\xi(t)$ is in the range $a\ll\xi\lr{(t)}\ll L$, see panel (\textbf{c}) in \cref{instaquench.Tc.plots}.
			
			Finally, the average sizes of the largest clusters evolve in a very similar way to the one found in the $T=0$ quenches: the only perceptible differences are that the smallest clusters do not disappear (thanks to the thermal fluctuations), and the dynamics are slightly slower (since $z_{\text{c}}>z_{\text{d}}$).
			See panel (\textbf{d}) in \cref{instaquench.0.plots}.
	\section{Effects of a finite cooling rate}
	\label{sec.cooling}
		\begin{figure}[!htb]
			\begin{center}
				\includegraphics{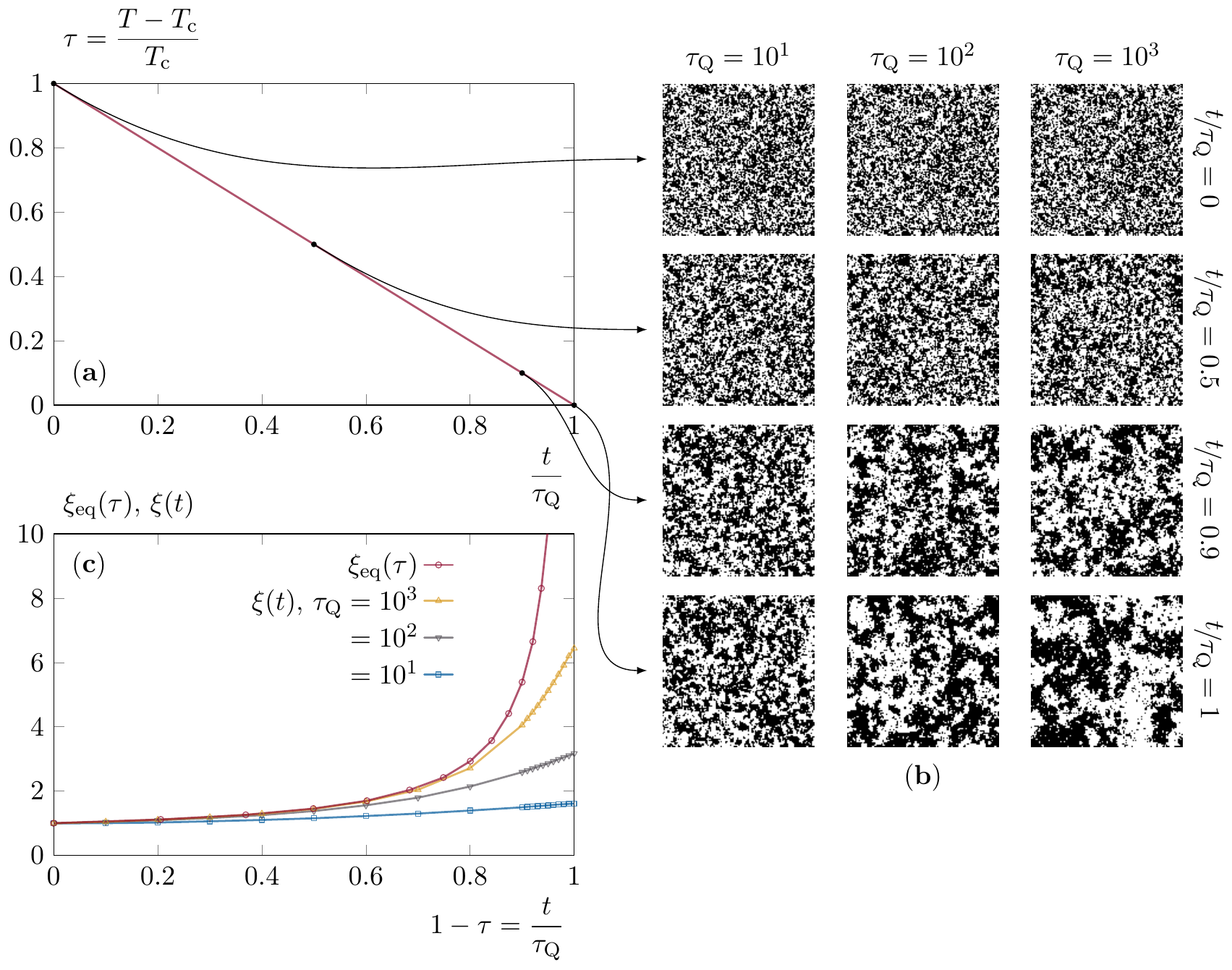}
			\end{center}
			\caption{
				(\textbf{a}) Description of the cooling process.
				Temperature is linearly decreases from $T=2\,T_{\text{c}}$ at $t=0$ to $T=T_{\text{c}}$ at $t=\tau_{\text{Q}}$.
				$\tau_{\text{Q}}$ controls the cooling rate, and the larger the values it takes, the slower the cooling.
				The right column (\textbf{b}) shows typical snapshots of the system ($L=128$) in the course of the cooling process, and for different values of the cooling rate.
				Panel (\textbf{c}) displays the evolution of the correlation length extracted from the space-time correlation function in the course of cooling in a system with $L=1024$ (note that the maximum value of $\xi$ is close to 10, much shorter than the system size).
				We have also represented the equilibrium correlation length at the corresponding temperatures.
			}
			\label{cooling.intro.plots}
		\end{figure}
		
		In the present section, we will discuss how the time spent in the vicinity of the Ising critical point affects the dynamics.
		
		Let us first describe the process considered in the remainder of this paper.
		The system is initially placed in an equilibrium state at $T=2\,T_{\text{c}}$ (\emph{ie} $T\lr{(t)}=2\,T_{\text{c}}$ for all $t\leq0$).
		Next, at $t=0$ the temperature of the bath is linearly cooled following
		\begin{equation*}
			\frac{T\lr{(t)}}{T_{\text{c}}}=2-\frac{t}{\tau_{\text{Q}}}\eqpc
		\end{equation*}
		where $\tau_{\text{Q}}$ is the cooling time up to the Curie temperature (see \cref{cooling.intro.plots}).
		In the present study, we only consider the dynamics above $T_{\text{c}}$ (\emph{ie} $t\in\lr{[0,\tau_{\text{Q}}]}$).
		Studies of the cooling rate effects on the coarsening dynamics that is at work close and below the critical point, even after annealing, have been presented in~\cite{Biroli10} for the $2d$ Ising model, in~\cite{Jelic11} for the $2d$ xy model, in~\cite{Priyanka} for a one-dimensional non-equilibrium lattice gas model with a phase transition between a fluid phase with homogeneously distributed particles and a jammed phase with a macroscopic hole cluster, and in~\cite{Comaron17a,Comaron17b} for time-dependent dissipative and stochastic Gross~-- Pitaievskii models relevant to describe micro-cavity polaritons and cold boson gases.
		\subsection{The \texorpdfstring{Kibble~-- Zurek}{Kibble - Zurek} mechanism}
			Starting from a thermal state, the system will follow the equilibrium conditions dictated by the changing environment as long as it can: \emph{ie} up to a time, called $\hat{t}$, when the time needed to thermalise becomes too long with respect to the relative rate of variation of temperature.
			Next, the system falls out-of-equilibrium and its further evolution will be discussed later.
			Obviously, the slower the cooling, the later the system will fall out-of-equilibrium.
			For an infinite system size, the time required to thermalise at the Ising critical point is infinite; it actually scales as $L^{z_{\text{c}}}$, and, unless cooling rates are scaled with the system size in a convenient way, the system will necessarily fall out-of-equilibrium at a certain point.
			Conversely, for finite-size systems, there exists a sufficiently slow cooling rate such that the system never goes out-of-equilibrium; we will discuss this point in \cref{sec.cooling.atTc}.
			We suppose the cooling to be sufficiently slow so that the system falls out-of-equilibrium only in a close vicinity of the critical point.
			On the one hand, in equilibrium, the correlation length depends on the distance from the critical point as $\tau^{-\nu}$.
			On the other hand, close to $T_{\text{c}}$, the dynamic correlation length grows in time as $\xi\lr{(t)}\sim{\lr{(t+\sharp\,{\xi_0}^{z_{\text{c}}})}}^{\nicefrac{\scriptstyle{1}}{\scriptstyle{z_{\text{c}}}}}$, where $\xi_0$ is the initial correlation length and $\sharp$ some constant factor.
			After an instantaneous quench from a state with correlation length $\xi(t=0)=\xi_0$ to a temperature at a distance $\tau>0$ from the critical point, the thermal state is reached after a time $t^{\text{th}}\lr{(\tau)}$ such that $\xi\lr{(t=t^{\text{th}}\lr{(\tau)})}\sim\xi_{\text{eq}}\lr{(\tau)}\sim\tau^{-\nu}$.
			Assuming that the instantaneous quench is performed from $2\,T_{\text{c}}$ and that the correlation length vanishes at this temperature, $\xi_0=0$, we have $t^{\text{th}}\lr{(\tau)}\sim\tau^{-\nu z_{\text{c}}}$.
			This is good estimate for the time needed to equilibrate at a distance $\tau$ from criticality.
			
			Now, following the argument proposed by Zurek~\cite{Zurek85}, the system falls out-of-equilibrium at a time $\hat t$, when the time needed to reach $T_{\text{c}}$, $\tau_{\text{Q}}-\hat{t}$ in the linear cooling procedure, becomes smaller than the time needed to thermalise at the current temperature $\hat T$ (the standard notation is such that the temperature and time at which the system falls out of equilibrium are noted by hats).
			Hence, we have
			\begin{equation*}
				\tau_{\text{Q}}-\hat{t}\sim t^{\text{th}}\lr{(\hat{\tau})}\sim{\hat{\tau}}^{-\nu z_{\text{c}}}\sim{\lr{(1-\frac{\hat{t}}{\tau_{\text{Q}}})}}^{-\nu z_{\text{c}}}\sim{\lr{(\frac{\tau_{\text{Q}}-\hat{t}}{\tau_{\text{Q}}})}}^{-\nu z_{\text{c}}}\eqpc
			\end{equation*}
			where $\hat{\tau}$ is the distance from the critical temperature at $\hat{t}$.
			Therefore, the system falls out-of-equilibrium at
			\begin{equation*}
				\hat{t}=\tau_{\text{Q}}-\sharp\,{\tau_{\text{Q}}}^{\nicefrac{\scriptstyle{\nu z_{\text{c}}}}{\scriptstyle{\lr{(1+\nu z_{\text{c}})}}}}\eqpc
			\end{equation*}
			where $\nicefrac{\displaystyle{\nu z_{\text{c}}}}{\displaystyle{\lr{(1+\nu z_{\text{c}})}}}\approx0.685$ and $\sharp$ another constant factor.
			
			In many papers dealing with the slow cooling of atomic systems the assumption is that, after $\hat t$, the system remains frozen and correlations do not grow beyond the correlation length present at this time,
			\begin{equation*}
				\hat{\xi}=\xi\lr{(\hat{t})}\sim\hat{\tau}^{-\nu}\sim{\lr{(\frac{{\tau_{\text{Q}}}^{\nicefrac{\scriptstyle{\nu z_{\text{c}}}}{\scriptstyle{\lr{(1+\nu z_{\text{c}})}}}}}{\tau_{\text{Q}}})}}^{-\nu}\sim{\tau_{\text{Q}}}^{\nicefrac{\scriptstyle{\nu}}{\scriptstyle{\lr{(1+\nu z_{\text{c}})}}}}\eqpc
			\end{equation*}
			where $\nicefrac{\displaystyle{\nu}}{\displaystyle{\lr{(1+\nu z_{\text{c}})}}}\approx0.315$.
			This, however, is not correct in coarsening systems as already discussed in~\cite{Biroli10,Jelic11,Priyanka,Comaron17a}, for example.
		\subsection{The out-of-equilibrium dynamics}
		\label{sec.cooling.atTc}
			At early times such that $t<\hat{t}$, the system evolves in equilibrium and the correlation length grows as the equilibrium one at the temperature reached at the measuring time:
			\begin{equation*}
				\xi_{{}_<}\lr{(t)}\sim\xi_{\text{eq}}\lr{(\tau)}\sim{\lr{(1-\frac{t}{\tau_{\text{Q}}})}}^{-\nu}\eqpd
			\end{equation*}
			When $t$ exceeds $\hat{t}$, the correlation length does not grow fast enough, and the system falls out of equilibrium.
			Now, since the equilibrium correlation length soon becomes much longer than the dynamic growing length, we can make the simplifying assumption that the system behaves as if it were in contact with a bath right at the Curie temperature.
			This proposal amounts to treating the problem as after an instantaneous quench at $t=\hat{t}$ from $\tau=\hat{\tau}$ to $\tau=0$.
			Hence, the correlation length continues to grow, but now as
			\begin{equation*}
				\xi_{{}_>}\lr{(t)}\sim{\lr{(t-\hat{t}+\sharp\,{\hat{\xi}}^{z_{\text{c}}})}}^{\nicefrac{\scriptstyle{1}}{\scriptstyle{z_{\text{c}}}}}\eqpc
			\end{equation*}
			where the term in ${\hat{\xi}}^{z_{\text{c}}}$ takes into account the non-vanishing correlation length at $t=\hat{t}$.
			
			Then, imposing the consistency of the correlation length before and after $\hat{t}$,
			\begin{equation}
				\xi_{{}_>}\lr{(\hat{t})}=\xi_{{}_<}\lr{(\hat{t})}=\hat{\xi}\sim{\tau_{\text{Q}}}^{\nicefrac{\scriptstyle{\nu}}{\scriptstyle{\lr{(1+\nu z_{\text{c}})}}}}\eqpc
				\label{eq.xidynamics.afterthat}
			\end{equation}
			we have
			\begin{equation*}
				\xi\lr{(t)}\sim\lr{\{\begin{array}{ll}
					{\lr{(1-\nicefrac{\displaystyle{t}}{\displaystyle{\tau_{\text{Q}}}})}}^{-\nu}&t\leq\hat{t}\\
					{\lr{(t-\tau_{\text{Q}}+\sharp\,{\tau_{\text{Q}}}^{\nicefrac{\scriptstyle{\nu z_{\text{c}}}}{\scriptstyle{\lr{(1+\nu z_{\text{c}})}}}})}}^{\nicefrac{\scriptstyle{1}}{\scriptstyle{z_{\text{c}}}}}&t\geq\hat{t}
				\end{array}.}\eqpd
			\end{equation*}
			where the second line is obtained from \cref{eq.xidynamics.afterthat} where we have replaced $\hat{t}=\tau_{\text{Q}}-\sharp\,{\tau_{\text{Q}}}^{\nicefrac{\scriptstyle{\nu z_{\text{c}}}}{\scriptstyle{\lr{(1+\nu z_{\text{c}})}}}}$ and $\hat{\xi}\sim{\tau_{\text{Q}}}^{\nicefrac{\scriptstyle{\nu}}{\scriptstyle{\lr{(1+\nu z_{\text{c}})}}}}$.
			In particular, when reaching the critical point,
			\begin{equation}
				\xi\lr{(t=\tau_{\text{Q}})}=\bar{\xi}\sim{\tau_{\text{Q}}}^{\nicefrac{\scriptstyle{\nu}}{\scriptstyle{\lr{(1+\nu z_{\text{c}})}}}}\sim\hat{\xi}\eqpd
				\label{cooling.Tc.xi.scaling}
			\end{equation}
			While Zurek assumes that the system is frozen immediately after falling out-of-equilibrium, here we claim that the dynamic growing length, $\xi\lr{(t)}$, continues to grow after $\hat{t}$.
			However, the growth between $\hat t$ and $\tau_{\text{Q}}$, when the cooling reaches $T_{\text{c}}$, only the pre-factor and not the scaling with $\tau_{\text{Q}}$ that is not modified.
			Therefore, if the interest is set upon the scaling properties of the system at the critical point (and not far below it) one can assume that the dynamic correlation length takes the form it had at $\hat{t}$.
			
			The next section will be devoted to putting eq.~(\ref{cooling.Tc.xi.scaling}) to the test.
			
			\begin{figure}[!htb]
				\begin{center}
					\includegraphics{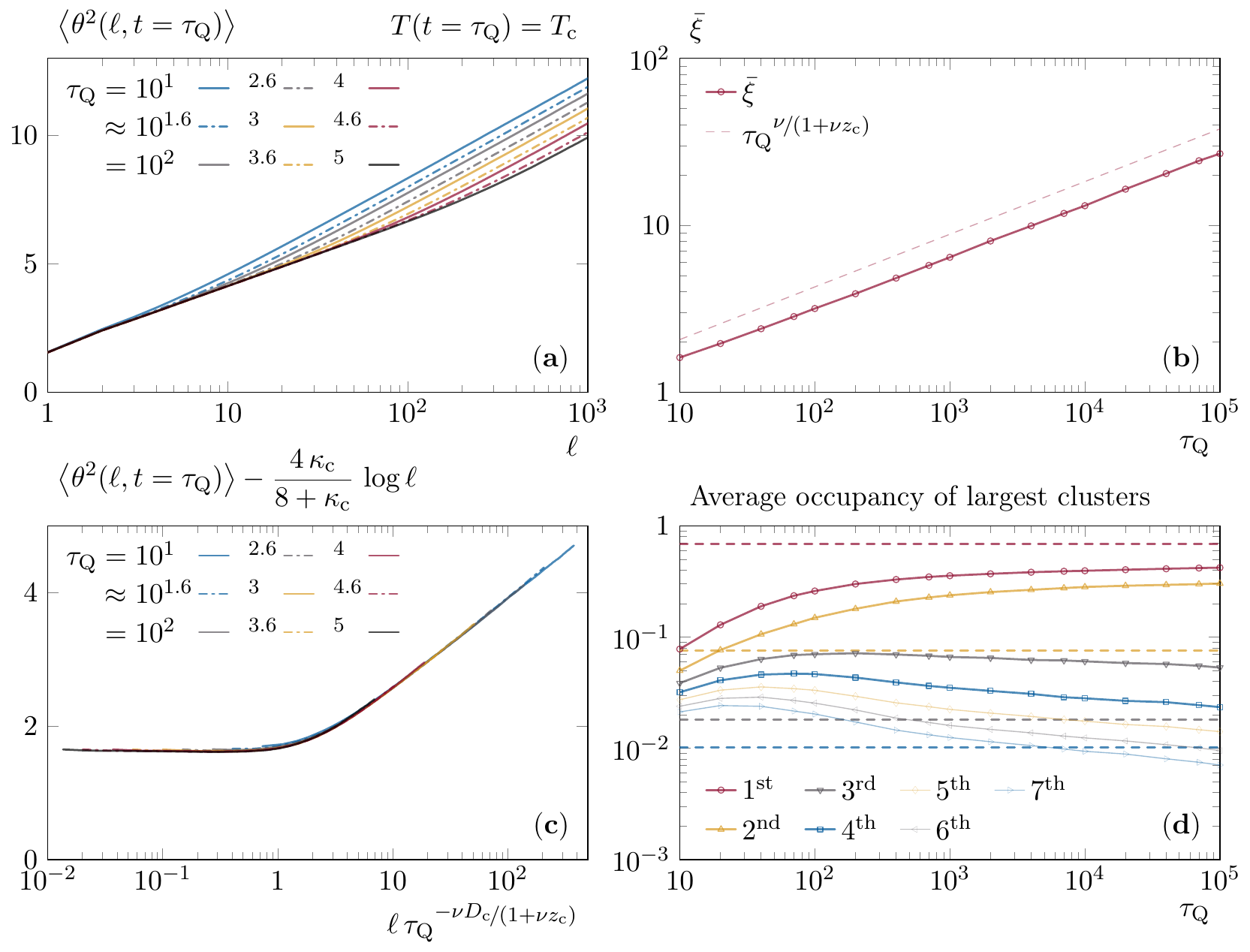}
				\end{center}
				\caption{
					Dependence on the cooling rate after a linear cooling to the critical point.
					Panel (\textbf{a}) represents the \textsc{wav} for different cooling times and panel (\textbf{c}) shows the same quantity, after the rescaling proposed in \cref{cooling.Tc.wav.scaling}; $\kappa_{\text{c}}=3$ and $D_{\text{c}}=1.375$ are the same as in \cref{sec.quench.atTc}.
					The graphics in (\textbf{b}) shows, as a function of the cooling rate, the measured correlation length when reaching the critical point; the dashed line is its predicted evolution (see \cref{cooling.Tc.xi.scaling}).
					The last figure, (\textbf{d}), represents the average occupancy rates of the largest clusters, and the dashed lines highlight the expected values for an infinitely slow annealing (\emph{ie} the values in equilibrium at $T_{\text{c}}$).
					All the results presented in these graphics were obtained with a system size of $L=1024$.
				}
				\label{cooling.Tc.plots}
			\end{figure}
			
			We are now going to describe the state of the system when reaching the critical point, and how it depends on the cooling rate.
			First of all, we can easily check that $\xi\lr{(\tau_{\text{Q}})}=\bar{\xi}\sim\hat\xi$ is quite an accurate prediction, see panel (\textbf{b}) in \cref{cooling.Tc.plots}.
			
			Next, let us analyse how the \textsc{wav} behaves: as exposed in \cref{sec.quench.atTc}, the interfaces present two critical properties: the Ising one on short length scales, and the percolation one otherwise.
			These features are proven in panel (\textbf{a}) in \cref{cooling.Tc.plots}.
			The length scale that separates the two behaviours scales with the effective correlation length when reaching the critical point.
			Thus, the rescaling
			\begin{equation}
				\lr{<\theta^2\lr{(\ell,t)}>}\to\lr{<\theta^2\lr{(\ell,t)}>}-\frac{4\,\kappa_{\text{c}}}{8+\kappa_{\text{c}}}\,\log{\ell}\quad\text{and}\quad\ell\to\frac{\ell}{{\bar{\xi}}^{D_{\text{c}}}}\to\ell\,{\tau_{\text{Q}}}^{\nicefrac{\scriptstyle{-\nu D_{\text{c}}}}{\scriptstyle{\lr{(1+\nu z_{\text{c}})}}}}\eqpc
				\label{cooling.Tc.wav.scaling}
			\end{equation}
			where $\nicefrac{\displaystyle{-\nu D_{\text{c}}}}{\displaystyle{\lr{(1+\nu z_{\text{c}})}}}\approx-0.434$, used in panel (\textbf{c}) in \cref{cooling.Tc.plots}, highlights the universal behaviour of the \textsc{wav}.
			The quality of this scaling provides a second proof of the accuracy of the prediction~(\ref{cooling.Tc.xi.scaling}).
			
			Consequently, when reaching the critical point, the system is (at least) thermalised up to a scale $s$, as soon as the cooling is slower than $\tau_{\text{Q}}^{{\text{th}}_s}$ which is such that
			\begin{equation*}
				s\sim\bar{\xi}\sim{\tau_{\text{Q}}^{{\text{th}}_s}}^{\nicefrac{\scriptstyle{\nu}}{\scriptstyle{\lr{(1+\nu z_{\text{c}})}}}}\quad\Rightarrow\quad\tau_{\text{Q}}^{{\text{th}}_s}\sim s^{z_{\text{c}}+\nicefrac{\scriptstyle{1}}{\scriptstyle{\nu}}}\eqpd
			\end{equation*}
			We recall that for an infinitely fast quench to $T=T_{\text{c}}$, the scale $s$ is thermalised after a time
			\begin{equation*}
				t^{{\text{th}}_s}\sim s^{z_{\text{c}}}\eqpd
			\end{equation*}
			Since $z_{\text{c}}+\nicefrac{\displaystyle{1}}{\displaystyle{\nu}}\approx3.17>2.17\approx z_{\text{c}}$, an instantaneous quench is more efficient than a linear cooling to create the structures of the Ising critical point; the time spent far from $T_{\text{c}}$ is not useful to develop the Ising criticality, the system develops, instead, the percolation one.
			
			This feature can also be observed by looking at the average sizes of the largest clusters by comparing panels (\textbf{d}) in \cref{instaquench.Tc.plots,cooling.Tc.plots}.
			Indeed, on \cref{instaquench.Tc.plots}, at $t=10^5\sim\nicefrac{\displaystyle{t^{{\text{th}}_{L=1024}}}}{\displaystyle{10}}$, the second largest cluster has already started to be swallowed by the first one, and the third and fourth have almost reached their equilibrium average sizes.
			In contrast, on \cref{cooling.Tc.plots}, at $\tau_{\text{Q}}=10^5$, all the largest clusters are still far from their equilibrium average sizes.
			Moreover, the first and second are still of the same order of magnitude.
			
			These results confirm that the dynamics are affected by the Ising critical point only in its close vicinity, and the time spent far from it is not helpful to get closer to equilibrium at $T_{\text{c}}$.
		\subsection{Dynamics before reaching the critical point}
		\label{sec.cooling.dynamics}
			\begin{figure}[!htb]
				\begin{center}
					\includegraphics{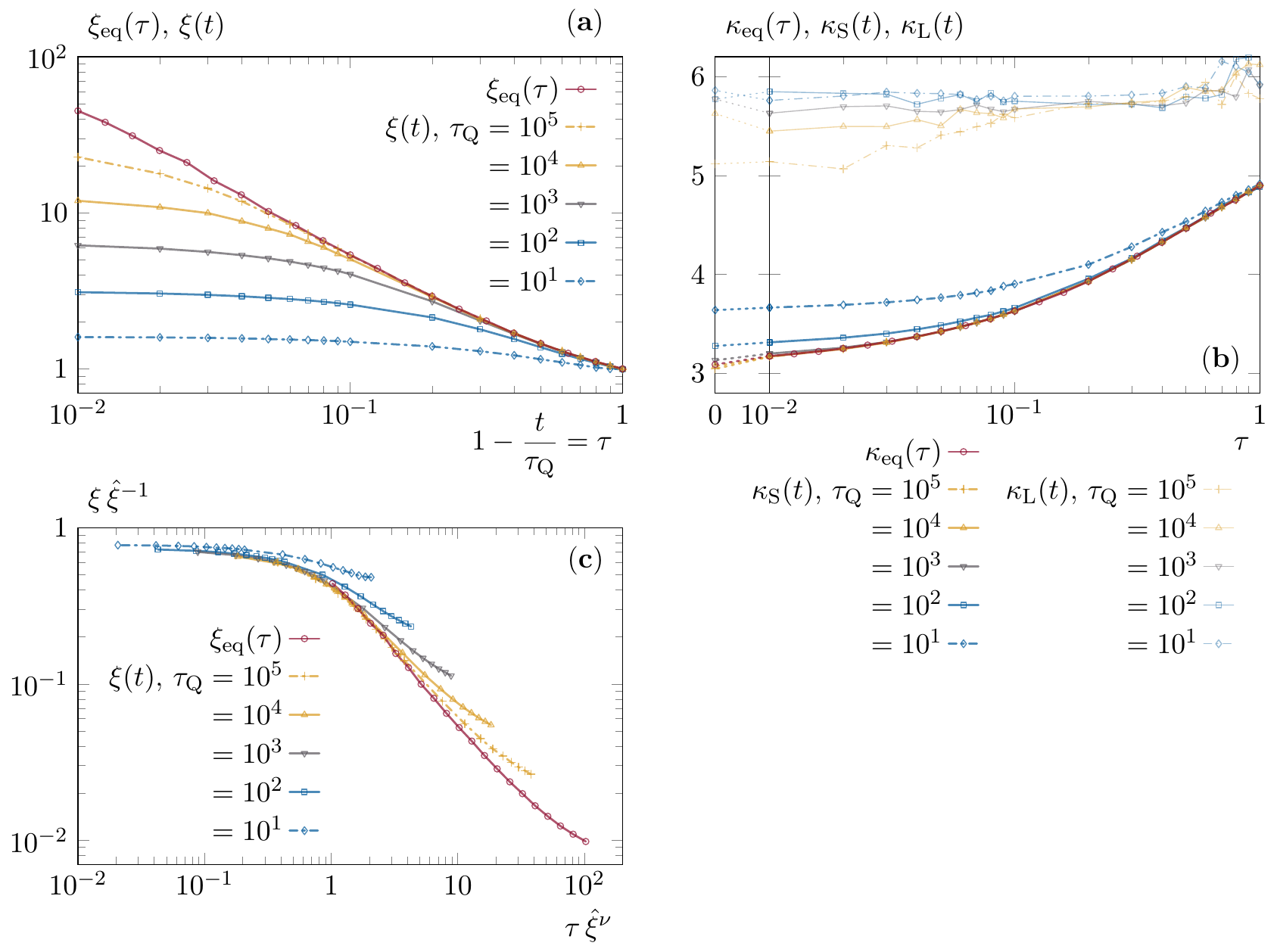}
				\end{center}
				\caption{
					Approach to the critical point; dependency on the cooling rate.
					Panel (\textbf{a}) shows the increase of the correlation length during cooling for different cooling rates; the equilibrium correlation length is also shown.
					Panel (\textbf{c}) represents the same quantities, but with a different scaling (following \cref{cooling.dyna.xi.scaling}).
					In panel (\textbf{b}), we show the evolution of the slopes of the \textsc{wav} when approaching the critical point together with the equilibrium one.
					$\kappa_{\text{S}}$ is extracted from the slope of the \textsc{wav} at short curvilinear length scales and is expected to have the Ising criticality when reaching the Curie temperature.
					$\kappa_{\text{L}}$ is extracted from the slope of the \textsc{wav} at long curvilinear length scales and corresponds to the percolation criticality.
					All the results presented in these graphics were obtained with a system size of $L=1024$.
				}
				\label{cooling.time.plots}
			\end{figure}
			
			In the previous section, we have shown that the behaviour when reaching the critical point does not really rely on the exact out-of-equilibrium dynamics in the range $t\in\lr{[\hat{t},\tau_{\text{Q}}]}$; whether the system remains frozen or evolves like in a post-quench dynamics.
			In this last section we try to clarify the situation.
			
			Let us consider that the system's typical length continues to grow as after an instantaneous quench after $\hat{t}$; the correlation length should then grow as
			\begin{align*}
				\xi\lr{(t)}&\sim{\lr{(t-\tau_{\text{Q}}+\sharp\,{\tau_{\text{Q}}}^{\nicefrac{\scriptstyle{\nu z_{\text{c}}}}{\scriptstyle{\lr{(1+\nu z_{\text{c}})}}}})}}^{\nicefrac{\scriptstyle{1}}{\scriptstyle{z_{\text{c}}}}}\\
				&\sim{\tau_{\text{Q}}}^{\nicefrac{\scriptstyle{1}}{\scriptstyle{z_{\text{c}}}}}\,{\lr{(\sharp\,{\tau_{\text{Q}}}^{\nicefrac{\scriptstyle{-1}}{\scriptstyle{\lr{(1+\nu z_{\text{c}})}}}}-\tau)}}^{\nicefrac{\scriptstyle{1}}{\scriptstyle{z_{\text{c}}}}}\eqpc
			\end{align*}
			and forgetting the dependence in the cooling rate, as
			\begin{equation}
				\xi\lr{(t)}\sim{\lr{(\sharp-\tau)}}^{\nicefrac{\scriptstyle{1}}{\scriptstyle{z_{\text{c}}}}}\eqpc
				\label{cooling.time.xi.tau}
			\end{equation}
			where the $\sharp$ factor has changed but is still positive.
			Thus, for $\sharp\gg\tau$ (or $\tau$ small enough), the correlation length is almost constant, and the system seems to be frozen.
			Moreover, the shape described by \cref{cooling.time.xi.tau} is in a quite good agreement with the numerical results presented in \cref{cooling.time.plots}(\textbf{a}).
			
			Let us now recall that the correlation length at the time or temperature at which the system falls out-of-equilibrium scales as
			\begin{equation*}
				\hat{\xi}\sim{\tau_{\text{Q}}}^{\nicefrac{\scriptstyle{\nu}}{\scriptstyle{\lr{(1+\nu z_{\text{c}})}}}}\eqpd
			\end{equation*}
			This is only valid while $\tau_{\text{Q}}$ is such that $\hat{\xi}\leq\xi_{\text{eq}}\lr{(\tau=0)}$.
			Beyond this point, the correlation length saturates to $\hat{\xi}=\xi_{\text{eq}}\lr{(\tau=0)}$; especially for an infinitely slow cooling (equilibrium).
			Doing the rescaling
			\begin{equation}
				\xi\lr{(t)}\to\frac{\xi\lr{(t)}}{\hat{\xi}}\quad\text{and}\quad\tau\to\tau\,{\hat{\xi}}^\nu
				\label{cooling.dyna.xi.scaling}
			\end{equation}
			(where $\hat{\xi}$ is taken as its saturation value for the equilibrium curve), the panel (\textbf{c}) in \cref{cooling.time.plots} shows that the correlation length has a universal behaviour.
			This is in agreement with a power law growth of the correlation length, as assumed in \cref{cooling.time.xi.tau}.
			Nonetheless, universality disappears far from the critical point since the equilibrium correlation length is subject to non-algebraic corrections in this region.
			
			Let us finally discuss the ``change in criticality'', as measured by the evolution of the parameter $\kappa$ in the course of the cooling process and compare it to the equilibrium ($\kappa_{\text{eq}}$).
			As done before, $\kappa$ is extracted from the \textsc{wav}.
			\Cref{cooling.time.plots} (\textbf{b}) represents the evolution of the criticality on short ($\kappa_{\text{S}}$) and long ($\kappa_{\text{L}}$) curvilinear abscissa length scales.
			The long length scales have almost always the criticality of percolation ($\kappa=6$) except for very slow cooling rates and in the vicinity of the Curie temperature where the criticality starts to be affected by the Ising critical point.
			On short length scales, the system is able to achieve the equilibration process, and the criticality corresponds to the equilibrium one discussed in \cref{sec:equilibrium} and represented by $\kappa_{\text{eq}}$.
			However, for the fastest cooling rates, \emph{eg} $\tau_{\text{Q}}=10^1$, even the short scales cannot follow the equilibrium.
	\section*{Conclusions}
		The purpose of this work was to study the influence of the cooling rate on the dynamics close to a second order critical point (between a symmetric and symmetry broken phase; here, for Ising models, the $\mathbb{Z}_2$ symmetry).
		More precisely, we analysed the evolution of the geometric and scaling properties of the interfaces between domain walls close and at the critical point.
		
		In order to set the stage, we first studied the fractal properties of the interfaces in equilibrium at various temperatures in the disordered phase.
		The analysis of the \textsc{wav} allowed us to reach our first conclusion:
		\begin{itemize}
			\item In equilibrium at $T>T_{\text{c}}$ the long-scale properties of the interfaces are the ones of critical percolation until a crossover length-scale that decreases with increasing temperature.
			A temperature dependent crossover towards critical Ising fractality at short-length scales arises close to the critical point, visible only below, say, $T=1.1\,T_{\text{c}}$.
		\end{itemize}
		
		So far, the influence of critical percolation on the dynamics of the $2d$ Ising~-- Glauber model after \emph{instantaneous} quenches from infinite temperature to the critical point~\cite{BlCuPi12} and below it~\cite{ArBrCuSi07,SiArBrCu07,BlCoCuPi14,BlCuPiTa17} was studied.
		The equilibrium result just mentioned indicates that this critical percolation geometry is present in high temperature equilibrium configurations.
		
	 	Next, we recalled some basic features of the coarsening dynamics following an instantaneous quench from equilibrium at $T= 2\,T_{\text{c}}$ both to zero and the Curie temperatures.
	 	On the one hand, we confirmed that correlation functions scale with a growing length that increases algebraically with time.
	 	On the other hand, we highlighted the non-trivial evolution of the geometry of the domains of parallel spins.
		The critical percolation geometry of the interfaces present in the initial state is progressively transformed, starting from the short scales, towards the one of the target temperature: smooth at zero temperature and critical Ising at $T_{\text{c}}$.
		
		We then explained the Kibble~-- Zurek mechanism~\cite{Kibble76,Zurek85} allowing one to estimate when the system falls out-of-equilibrium while approaching a critical point from the symmetric phase with a finite speed.
		While Zurek assumes that the system freezes after falling out-of-equilibrium, following~\cite{Biroli10,Jelic11,Comaron17b} we argued that the correlation length continues to grow in this regime as if the system were instantaneously quenched to the critical point.
		Our argument does not affect the scaling in the cooling rate predicted by Zurek, but offers a more accurate description of the growing of the correlation length after the system has fallen out-of-equilibrium.
		We examined this scaling numerically and we found that 
		\begin{itemize}
			\item after a slow linear cooling with rate $\tau_{\text{Q}}$ to $T_{\text{c}}$, the dynamic growing length extracted from the analysis of the space-time correlation function scales as $\xi(\tau_{\text{Q}})\sim\tau_{\text{Q}}^{\nicefrac{\scriptstyle{\nu}}{\scriptstyle{\lr{(1+\nu z_{\text{c}})}}}}$.
		\end{itemize}
		
		During the slow cooling process, while far from the critical point, the interfaces keep the fractal properties of critical percolation over a wide range of length scales, up to a temperature dependent crossover length.
		However, when approaching temperatures close enough to the critical point, we observe that 
		\begin{itemize}
			\item the winding angle variance satisfies a scaling with respect to $\xi(\tau_{\text{Q}})$, and besides, the interfaces with critical Ising properties span the length scales that are shorter than ${\tau_{\text{Q}}}^{\nicefrac{\scriptstyle{\nu D_{\text{c}}}}{\scriptstyle{\lr{(1+\nu z_{\text{c}})}}}}$.
		\end{itemize}
		
		Finally, our results prove that the Ising critical point influences the dynamics only in its close vicinity.
		Therefore,
		\begin{itemize}
			\item an instantaneous quench procedure is much more efficient to create the structures of the Curie temperature than a slow annealing.
		\end{itemize}
		As a matter of fact, the time spent far from the Ising critical point does not contribute to the thermalisation of the system; instead, the dynamics of the system is governed by critical percolation.
		
		This study is also a complement to works that try to elucidate the role played by the initial conditions on the post-quench dynamics of the Ginzburg~-- Landau scalar field theory~\cite{BrayHumayunNewman} and, more recently, the kinetic Ising model~\cite{ChakrabortyDas15,ChakrabortyDas16,Corberi16} as well as the influence of a non-vanishing cooling rate on the scaling properties of discrete models close to their phase transition~\cite{Liu-etal14}.
		In the latter paper the emphasis was set on the scaling properties of the order parameter and how these depend, or not, on the microscopic stochastic updates.
		We focus instead on the geometrical and scaling properties of the structures when slowly approaching the critical point.
	\paragraph{Acknowledgements.}
		L.~F.~C. is a member of Institut Universitaire de France.
		We thank A. Tartaglia for very useful discussions and M. Henkel for his helpful comments on the manuscript.
	\bibliographystyle{phaip}
	\bibliography{coarsening}
\end{document}